# Fostering continuous innovation in design with an integrated knowledge management approach

Jing Xu [a, b, c]*, Rémy Houssin [a, b], Emannuel Caillaud [a, b], Mickaël Gardoni [b, d]

[a] Université de Strasbourg, UFR Physique et Ingénierie, 25 Rue Maréchal Lefbre, 67100, Strasbourg, France ;
[b] INSA Strasbourg – LGECO, 24 Boulevard de la Victoire, 67084 Strasbourg Cedex, France ;
[c] Yangzhou University, College of Mechanical Engineering, 196 Huayang West Road, 225127, Yangzhou, China;
[d] Université du Québec, Ecole de Technologie Supérieure, LIPPS, 1100 rue Notre Dame Ouest, Montréal, H3C 1K3(Canada);

Abstract
In the global competition, companies are propelled by an immense pressure to innovate. The trend to produce more new knowledge-intensive products or services and the rapid progress of information technologies arouse huge interest on knowledge management for innovation. However the strategy of knowledge management is not widely adopted for innovation in industries due to a lack of an effective approach of their integration. This study aims to help the designers to innovate more efficiently based on an integrated approach of knowledge management. Based on this integrated approach, a prototype of distributed knowledge management system for innovation is developed. An industrial application is presented and its initial results indicate the applicability of the approach and the prototype in practice.

* Corresponding author. Tel.: +33 388 14 47 00; fax: +33 388 24 14 90.
E-mail addresses: jing.xu; remy.houssin; emmanuel.caillaud, gardoni @insa-strasbourg.fr

## 1. Introduction

Under the intensive competition of the global market, companies are urged to innovate in order to succeed even survive. It is reported that successful companies produce 75 percent of revenues from new products or services that did not exist five years ago [1]. Thanks to the trend of more knowledge-intensive products or services, the competition based on knowledge and innovation is highly valued by companies. While how to turn the available knowledge into future innovations in a continuous way is a research problem for both academics and industries. Recently Knowledge Management (KM) as an emergent discipline has gained much attention for this subject [2, 3]. During the passed decades, a great deal of research on KM has been carried out from the viewpoints of management and engineering. However, the KM strategy is not well adopted in industries due to the difficulties of managing knowledge for innovation. There is still a lack of an integration of the mechanisms of KM with innovation from a systems thinking perspective [4, 5].

Engineering design becomes a more knowledge-intensive process, which is defined as a structured creative process bringing new ideas and knowledge into use either as product (service) or process innovations [6]. Innovation has been regarded as an inherent nature of design process [7]. At the same time, the research on knowledge and its management emphasizes the importance of knowledge on innovation and design [7, 8]. Due to the complexity and emergent nature of design, various groups of knowledge should be provided at the right time and at the right place and be managed for reducing the complexity and uncertainty of innovation [9].

Managing knowledge for innovation has difficulties in the reconcilement of diverse perspectives of KM and innovation, the distribution of heterogeneous knowledge, and the balance between exploration and exploitation [10]. As the increasing amount of knowledge is embraced by innovation, it becomes a prerequisite to manage the knowledge and to leverage it into innovation in order to sustain the competitive advantage of a company. Thus, our main objective in the study is to foster the innovation in design and to help the designers innovate with an integrated approach of KM.

In section 2, related work about KM and innovation in literature is presented and the disadvantages of current extant KM and innovation supporting systems are discussed. As a result, there is a need of a new KM approach for supporting innovation. In section 3, an integrated approach of KM for innovation is presented by using the Unified Modelling Language (UML). After, section 4 concerns the design of the framework for an agent-based distributed KM system based on this integrated approach. In section 5, the detailed implementation of a software prototype is presented. Then an industrial application is illustrated in section 6 for exhibiting the applicability and effectiveness of our approach in practice. Finally, the contributions of our work are discussed and the future research is planed.

## 2. Related work

Design and innovation have many meanings from different perspectives, but in our study we focus on their engineering aspect. A crucial endeavour of companies is to both generate new ideas and knowledge and efficiently convert them into new products [11]. As the basis of design, engineering information and knowledge have been considered as the crucial assets for its efficiency and innovativeness. Due to the diversity and specialization of knowledge, engineers have different requirements about knowledge [12]. They also need various methods to manage their knowledge for innovation in design.

### 2.1. Complex interactions of KM and innovation

In engineering design, knowledge is the basis for reasoning and problem solving [7]. In a design project, knowledge is created and used for explicit aims and in a specific context. Due to its multifaceted nature, same knowledge may have different meanings for different persons. The importance of contextual information has been recognized for better understanding of knowledge [13]. It can lead to better decision making and problem solving in design. According to the various properties of knowledge in [14-16], knowledge should include both the static and the dynamic aspects. Thus, we adopt the working definition of knowledge by Davenport and Prusak [15] and argue that a systemic model of knowledge for innovation should encompass both aspects of knowledge regarding the characteristics of innovation such as novelty and appropriateness in [17].

In the knowledge-based economy, knowledge has been seen as one of the sustainable assets and companies have paid much interest on KM and its systems. Among the diverse perspectives of KM and innovation, the process approach is prevalent in engineering research. And it has specific advantages for design and innovation in avoiding information overload, enhancing value creation and improving knowledge usability [3].

On the one side, KM has evolved through several generations and a lot of KM models and processes have been proposed and used in companies with different emphases on human resources, company culture or information technologies. On the other side, innovation has been a long topic of research and various generations of innovation models have been put forward and applied in practice. According to our close investigation in [18] about existing KM and innovation models, the human, physical and technological perspectives of KM have been distinguished and knowledge creation and usage have been identified as two core KM activities for innovation. The models of KM and innovation have been compared in table 1 and more detailed discussions can be found in [18].

Table 1. Corresponding models of KM and innovation in [18]

| Models of Innovation | Models of Knowledge Creation Models of Knowledge Usage |
|---|---|
| Technology push model (1950s-1960s) | Knowledge production (1960s) Technological model (1960s) |
| Demand pull model (1960s-1970s) | Problem-solving methods (1970s) Economical model (1970s) |
| Coupling model (1970s-1980s) | Data-information-knowledge model (1980s) Institutional model (1980s) |
| Parallel and evolutionary model (1980s-1990s) | SECI Spiral, Knowledge Pentagram System (1990s) Social interaction model (1990s) |
| Innovation with systemic integration (1990s-2000s) | Nanatsudaki Model (2000s) Systems approach for knowledge use (2000s) |
| Open/Continuous model of Innovation (2000s) | *No correspondent model* |

In the table 1, we notice that different KM models match different generations of innovation models in literature except for the continuous model of innovation. Previous lessons learned have show that KM initiatives as a separated process in a company had almost failed to support innovation. Thus, an integrated approach of KM is required for continuous innovation process, which needs to synthesize diverse perspectives of KM from the view of systems thinking.

### 2.2. Current KM and innovation supporting systems

The rapid progress of advanced Information and Communication Technologies (ICT) has become an important enabler for KM and played an important role on successful KM initiatives. Different information technologies appear during the generations of KM [19]. Some are existed information tools borrowed from other disciplines for KM functions such as traditional information systems, document management systems, relational and object databases systems and data warehousing. Others are ICT tools developed for KM needs from their inception such as groupware, intranet, expert system, E-learning, help desk applications, collaboration and communication support system and so on. For example, Houssin et al. [20, 21] have proposed an information system for capitalizing security knowledge in design. Most of these systems are adapted for KM incentives, which mainly support to manage structured information, content and documents. Consequently, the diverse requirements of engineers, their behaviours and attitudes towards knowledge and KM are not well considered in them [12]. Meanwhile the contextual information about design is not fully supported by existing KM systems [13]. They also suffer several other limitations: the quality of system, and the incongruence and untrustworthiness of knowledge [22]. Moreover, most existing KM systems and tools focus on knowledge codification, sharing and reuse [2, 23] and few of them are intended to support innovation.

Recently, a new category of methods and tools emerges in the field of computer-aided innovation (CAI) as a response to the industrial demands for innovation [24], which are founded on diverse innovation theories such as TRIZ, axiomatic design, brainstorming, mind mapping and so on. On the one side, there exist many idea and innovation management systems based on the methods such as brainstorming and mind mapping etc. They stress the provision of right circumstance and methods for ideation and its implementation. They are largely created and used from the view of management without enough considerations of engineers' requirements [12]. Most of them do not distinguish ideas from knowledge for innovation and few adopt an explicit strategy of KM.

On the other side, some CAI tools are based on the systematic innovation methodologies especially such as TRIZ. They have difficulties to fully integrate with the design activities [11]. And such tools heavily depend on the technical and patent databases and require specific problem modelling techniques [25, 26]. They also endure other impediments such as the difficulty to use and interpret the general principles and laws in specific design situation and the poor interoperability of its results with other existing systems. This may account for the inadequate diffusion of such CAI tools in industry.

Due to the emergent nature of design, innovation in design can be seen as an exploration or expansion of the ill-defined design space [27, 28]. Knowledge as an essential asset plays an important role in innovation. Although there are urgent demands for innovation and eager wishes to utilize computer aided systems, KM strategies and CAI tools have not been widely adopted to foster innovation in design. That is partially due to the unavailability of a cogent approach of KM for innovation and the limitations of current computer support systems. In the following section, an integrated KM approach is presented as a base for innovation in design.

## 3. The integrated approach of KM for innovation

With increasing amount of knowledge involved in innovation and design, the successful management of engineering knowledge is an essential task for innovation in design. Due to the multidisciplinary nature of both KM and innovation, we did not find, in literature, any framework that generally accepted for linking KM with innovation. Thus from the systems thinking perspective, an integrated approach of KM is presented with a systemic model of knowledge and a hierarchical model including macro process and meta-model of KM. For a general understanding and readability of the models, the integrated approach of KM is presented by using the Unified Modelling Language (UML) techniques [26, 29].

### 3.1. A systemic model of knowledge for innovation

Based on the formulations of C-K theory [27], design as the generation of new object can be modelled by the co-expansion of concept and knowledge space. Knowledge as the crucial asset for innovation evolves in an ill-defined design space. It is increasingly recognized that not only the content of knowledge but also its context are of great importance for its better understanding and application in decisions and actions [12, 13, 15]. In the same way for the traceability and trustworthiness of knowledge, both static and dynamic aspects of knowledge should be included. By integrating a general framework of context in activity theory [30] with the relevant dimensions of knowledge attributes in [15], a systemic model of knowledge is built by UML in figure 1.

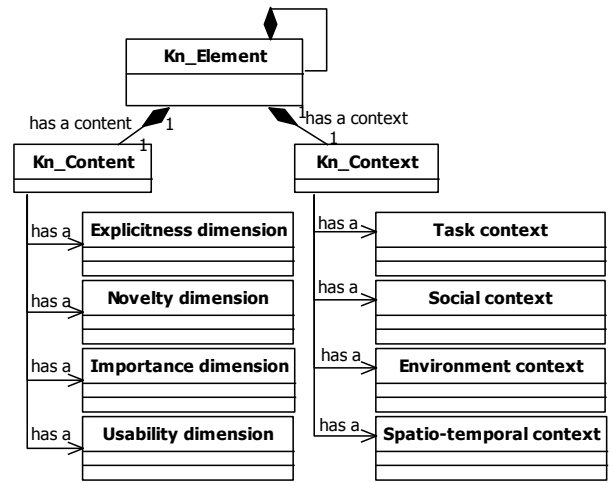

Figure 1 The systemic model of knowledge

In the model, each knowledge element (Kn_Element) is composed of its content (Kn_Content) and its context (Kn_Context). Kn_Content concerns the static aspects of knowledge with four dimensions that are extracted in terms of innovation characteristics such as explicitness, novelty, importance and usability. Each dimension is constructed as a continuum with two extremes at the ends and several possible locations in between [16]. For example, the Explicitness dimension has five levels with integer values from 1 to 5: totally tacit, more tacit than explicit, semi-tacit and explicit, more explicit than tacit and totally explicit. The same structure has been applied to the other three dimensions. Thus, all of the four dimensions have values and semantics that can indicate the quality of a piece of knowledge.

As to the context model of knowledge, Kn_Context is composed of four sub-contexts respectively relating to the context of creation and usage of knowledge. The sub-contexts concern engineering design situations such as the design tasks, places, design teams and other resources used in design process. They capture where, when, how and by who the knowledge is created and used. With the help of agent technology, the context can be automatically collected and stored in the knowledge base.

As knowledge evolves in organizations, the values and semantics in the dimensions of its content and its context also change according to our understanding about it. The mechanisms of the version control and the parent-child relationship of knowledge elements are used for tagging their evolutions. And they can be depicted by knowledge flows and networks that are visualized as a graph with nodes and links. With the inclusion of evolving content and contextual information in the knowledge base, the traceability and reliability of knowledge elements can be greatly enhanced and this lead to better insights and decision making in design. Later, a computational model of knowledge element is constructed based on the systemic model in a KM system.

3.2. The hierarchical model of KM for innovation

Knowledge has been regarded as the essential asset for the competitive advantage and innovation is vital to sustain it. For innovation to occur, knowledge needs not only to be created and shared, but also to be used and recombined. Innovation is the application of knowledge to produce new knowledge [31]. Since innovation is a result of the creative combination of existing knowledge [32], the seamless integration of knowledge creation and usage is crucial for the continuous innovation. In practice, innovation is commonly defined as ideas successfully applied in practice, but here we concentrate on exploring how ideas can be turned into successful products or services in the process of innovation from the KM perspective. Thus from this view, we argue that innovation can be seen as the application and recombination of existing and new knowledge to create more new knowledge embedded in it.

By extending the KM framework proposed by Alavi and Leinder in [33], an extended lifecycle of knowledge is constructed with focusing on the two core activities in innovation: knowledge creation and knowledge usage. Five phases are included in the new knowledge lifecycle model such as Pre-creation, Creation, Intermediate, Usage and Post-usage. A macro process of KM based on the knowledge lifecycle is modelled in UML as a state chart diagram of knowledge element in the figure 2.

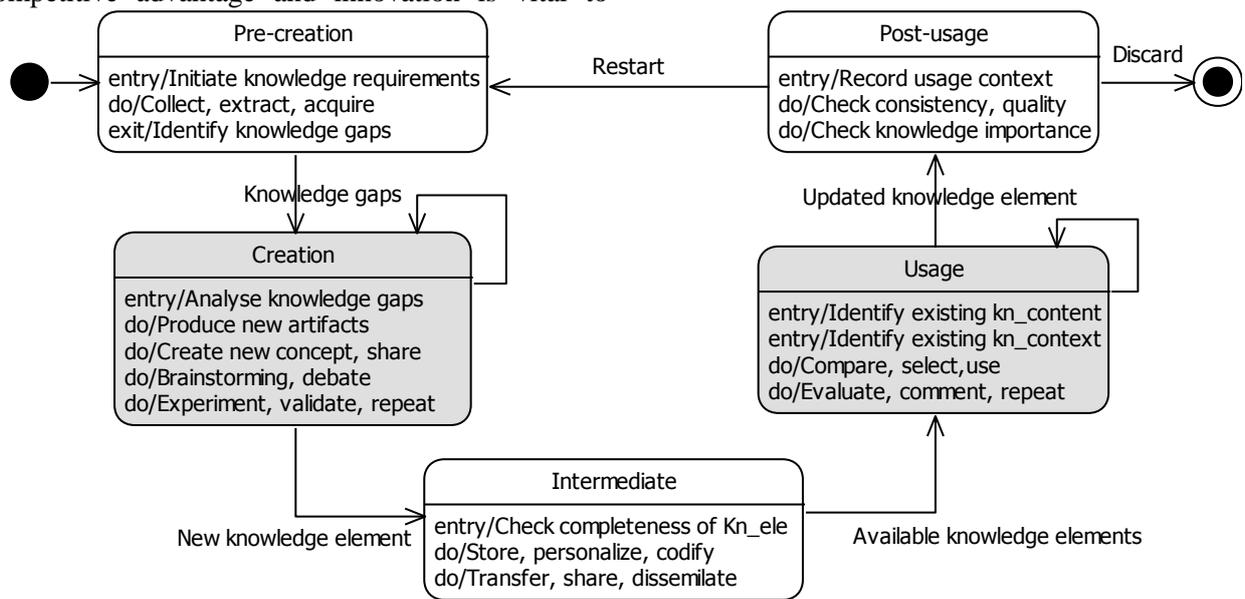

Figure 2 Macro process of KM for innovation

In each phase of the lifecycle, the macro process includes various KM activities, which could be executed in a concurrent and collaborative manner. Due to the distributed nature of knowledge and the limitations of time and space, the creation and usage of knowledge scatter unevenly in a company. As innovation requires both to create new knowledge and to use existing knowledge efficiently and effectively, the seamless integration of both knowledge creation and usage is imperative for it. Thus, under the networking and systems integration of the two activities, new knowledge emerges from their intensive interactions. This is explicated by a meta-model of KM in figure 3.

In the meta-model of KM, knowledge creation and usage contain two aspects with reference to innovation: one for innovation and the other for non-innovative tasks. True creation and Creative use focus on creating really new knowledge and creatively using it to solve innovative tasks. While Self learning and Routine use aim for improving personal knowledge repository and applying extant knowledge for non-innovative activities. The meta-model illustrates an ideal state for KM where the intensive networking and system integration diminish the gaps between the creation and usage of knowledge. At the same time they are supported by other activities in the macro process of KM with the help of ICT tools.

At a micro and individual level, the meta-model of KM describes the crucial knowledge activities of individuals for innovation. Whereas in the macro and organizational level, the macro process of KM explains the management of organizational knowledge in a company. On the one hand, the meta-model highlights the importance of human creativity and heterogeneity of knowledge for the novelty in innovation. On the other hand, the macro process stresses the integrity and lifecycle of KM in the total innovation process. For accelerating the flow of knowledge, the meta-model and macro-process of KM collaborate and interact with each other. Their dynamics are illustrated in a hierarchical model with four layers in the figure 4.

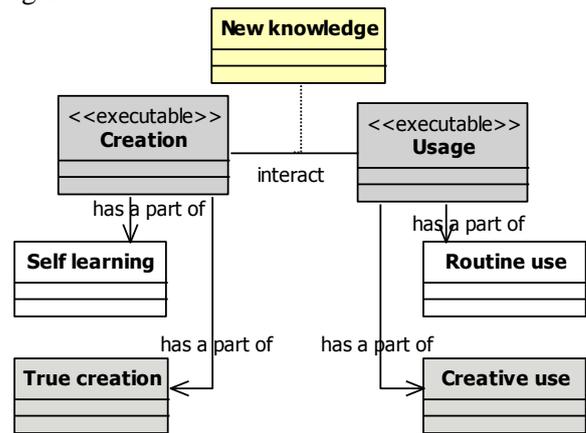

Figure 3 Meta-model of KM for innovation

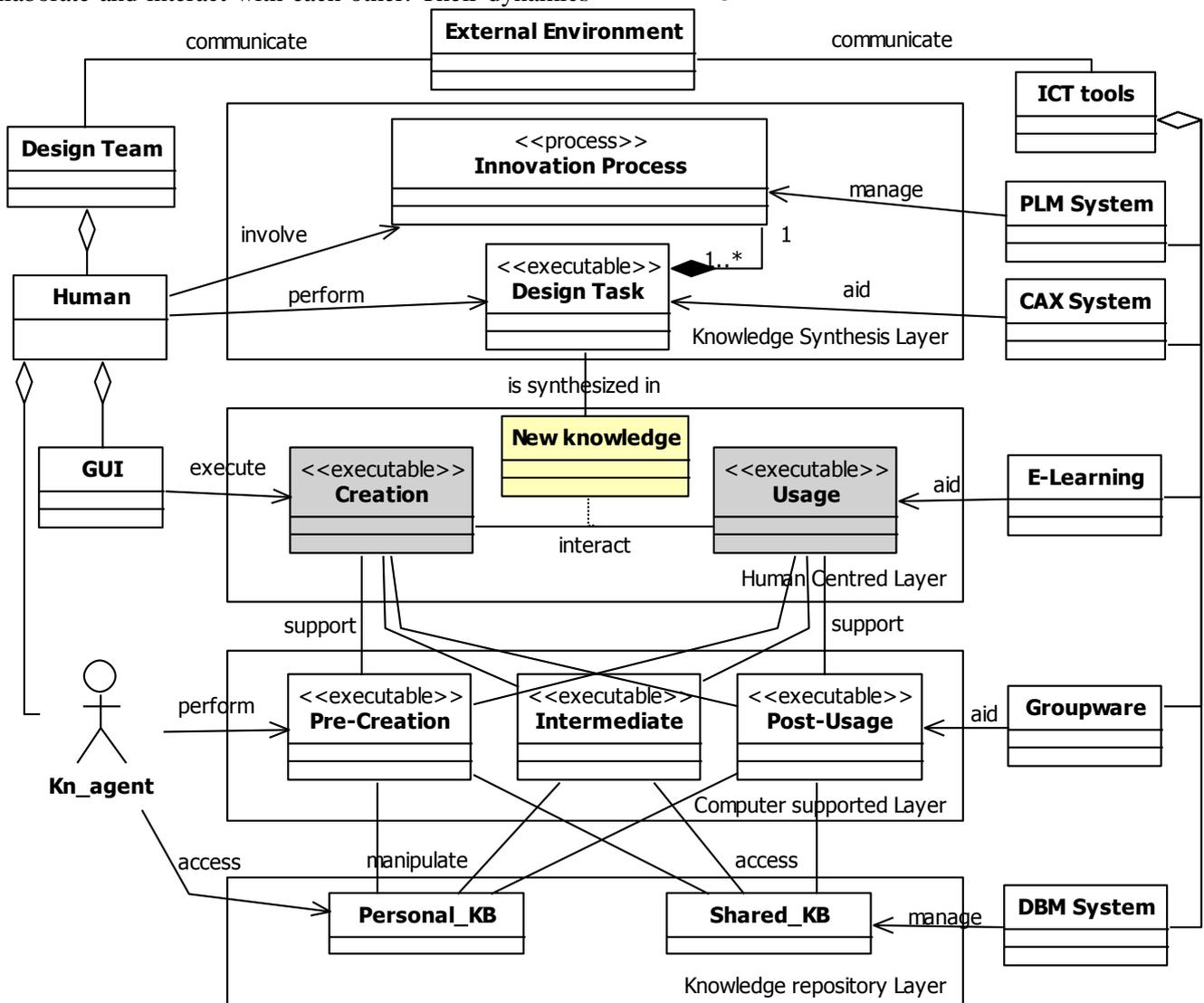

Figure 4 The hierarchical model for KM and innovation

Four layers are distinguished according to the sequences of knowledge flow in innovation. They are knowledge repository layer, computer supported layer, human centred layer and knowledge synthesis layer. Each layer embodies different roles of KM for innovation. On the knowledge repository layer, the personal knowledge bases (Personal_KB) and the shared knowledge base (Shared_KB) provide a solid basis for innovation and they are managed by database management system (DBM System). On the computer-supported layer, the

Pre-creation, Intermediate and Post-usage phases are arranged to support the creation and usage of knowledge. Here knowledge is processed by advanced ICT tools and pushed into the upper layer for the creative abrasion. On the human centred layer, the phases of Creation and Usage are placed thanks to the intensive human interventions in them. Their interactions lead to emergence of new knowledge that is synthesized in innovation. Then, Innovation Process and Design Task are situated on the knowledge synthesis layer, where new and existing knowledge are synthesized and implemented in innovation. Finally, the knowledge created and used in innovation flows back to the knowledge repository layer and restart a new cycle of knowledge flow.

In the hierarchical model, a multi-functional Design Team and ICT tools are incorporated as two important enablers for communicating and collaborating with the External Environment. The multi-functional Design Team consists of different stakeholders coming from various domains. And they provide innovation with distributed knowledge from internal or external sources. As the availability of ICT tools has been instrumental in catalyzing KM and innovation [14], they can help to relieve human efforts to process knowledge in each layer. With the two enablers, the knowledge flow among the four layers is accelerated and the availability of the knowledge for innovation has been improved.

In a word, the integrated approach of KM for innovation presented here in UML contains the systemic model of knowledge and the hierarchical model for leveraging KM activities into innovation process. The integration of KM into the innovation process makes KM no longer a separated process in a company, and offers a method to examine its business activities from the viewpoint of KM. With the multifunctional design team and advanced ICT tools, the gaps between knowledge creation and usage are significantly bridged through networking and system integration. This integrated approach can respond to the requirements of the continuous model of innovation and relieve some limitations of current computer support systems to a certain extent. In the following section, a distributed KM system framework will be designed in detail based on the approach.

# 4. KM system framework based on the integrated approach of KM

This section first defines the objective and functionalities of a KM system based on the integrated approach of KM. And then, according to the hierarchical model, an agent-based and distributed framework of the KM system is designed. Finally, the system is modelled with the static and the dynamic diagrams in UML.

## 4.1. Objective of the KM system

Developing a KM system for innovation is a complicated task that contains not only the technical issues but also people's concerns about KM. As Product Lifecycle Management (PLM) tools are being integrated with the KM methods and tools, new alternatives of computer supported innovation tools arise for the creation of new engineering desktop paradigms [24]. Thus, the objective of our system is identified as "to help designers innovate more easily and efficiently through focusing on the creation and usage of engineering knowledge based on the integrated KM approach".

## 4.2. Function requirements of the KM system

As the emergent nature of design, innovation can be seen as the exploration and expansion of design space [27, 28], which are characterized by the knowledge creation and usage. In order to foster innovation with the integrated approach of KM, the distributed KM system should include the following functional components for example the functions of knowledge creation, knowledge usage, task and agent management etc.

Different with existing KM systems, our system first focuses on the functions of creation and usage of knowledge that directly create values for innovation. Designers can create new knowledge through fulfilling various design activities such as self-learning, reflection and practice. When knowledge is to be used, it is critical to identify the user's current context and verify the appropriateness of the usage of such knowledge. After its usage, the user's context is automatically recorded in knowledge base for its traceability and trustworthiness. At the same time, users are encouraged to evaluate and rank the value of knowledge elements according to the usage.

Meanwhile, with the available ICT tools, our KM system also provide a common support for other KM activities in the phases of pre-creation, intermediate and post-usage. As illustrated in the hierarchical model, traditional information system such as groupware and database management system etc. can support the functions such as storage, retrieval, and navigation of knowledge. Because of available existing ICT tools, most of them are integrated as the background functions in our system.

Besides the above KM functional requirements, other functions are also considered such as the functions to be compatible and collaborate with existing information infrastructure: the PLM system and CAD/CAE systems. As agent technology is used to assist users, the management of various agents and their communications and collaborations are another necessary function of our system. All these identified functions are incorporated into the system framework and are implemented by different systems modules as shown in the table 2.

Table 2 Functional requirements with system interfaces

| | Functional requirement | Relevant system interfaces |
|---|---|---|
| KM Function | Knowledge creation | Knowledge Audit Interface |
| | Knowledge usage | Knowledge Network Interface |
| | Supporting KM activities | DBM system and XML files organization |
| System Management | Task management | PLM Interoperation Interface |
| | User management | User Management Interface |
| | Agent management | Agent Management Interface |

4.3. System framework design

The system framework is designed according to the functionalities of our system. As we discussed above, the hierarchical model for KM and innovation can work as an individual working framework in the KM system. Since no individual can have the sufficient knowledge and capability to fulfil a whole innovation, it is necessary to involve various people in an innovation project. They are organized into a multi-functional team and each member collaborates with one another. The system framework is designed by adopting the agent paradigm as shown in figure 5.

On the design platform, innovation design project is decomposed into design tasks that are assigned to the members of a design team. They collaborate to fulfil the assigned design tasks. Each member creates and uses knowledge through a graphical user interface with the support from his personal knowledge agent. The personal knowledge agent provides KM support activities autonomously and communicates with other agents in the agent community according to the requirements coming from its owner. The personal knowledge agent accesses the personal and shared knowledge bases for storing and retrieving relevant knowledge.

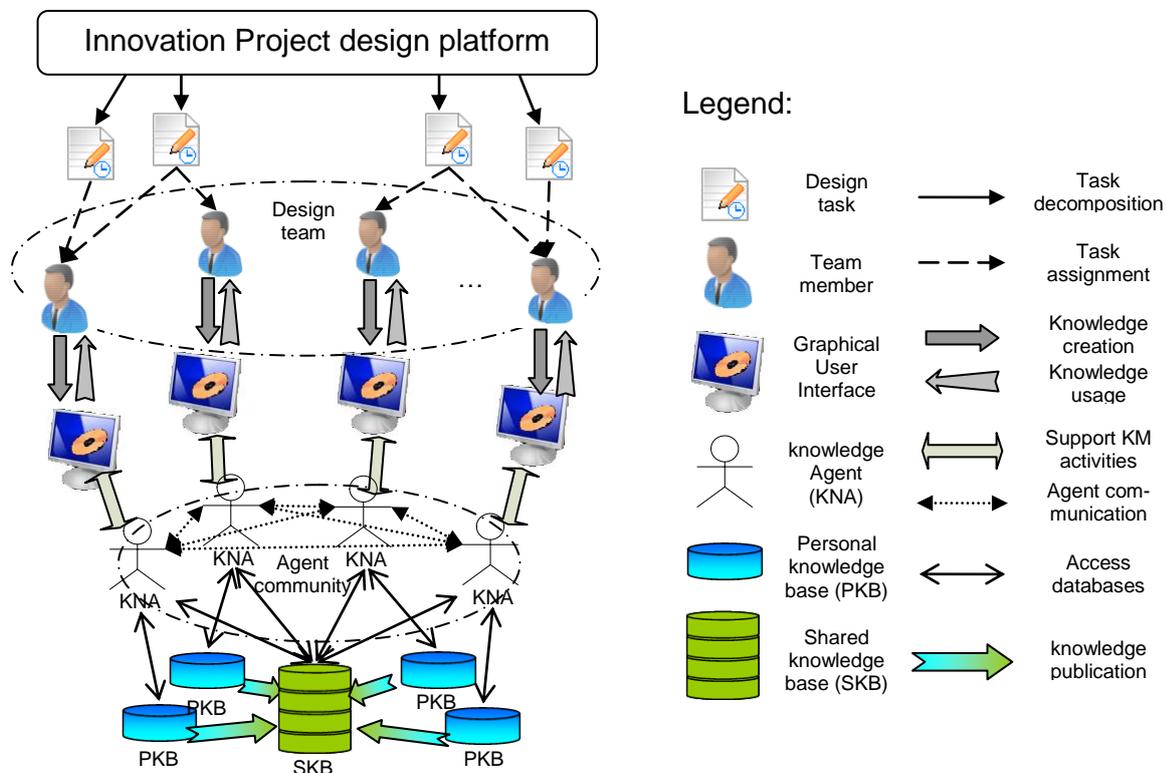

Figure 5 Framework of distributed KM system for innovation

This system framework is a detailed implementation of our hierarchical model for KM and innovation. The innovation project design platform and design tasks correspond to the knowledge synthesis layer in the hierarchical model. The multidisciplinary Design team and the Graphical User Interface (GUI) constitute the human centred layer where knowledge creation and usage are performed by humans. The agent community together with the GUI embodies the computer support layer where supporting KM activities are located and accelerated through ICT tools. The personal and shared knowledge bases form the knowledge repository layer accessed by agent community. In the following, the KM system is modelled in detail in relation to the system framework.

4.4. System modelling in UML

Under the system framework, system modelling aims to define the generic computational models in detail for the system realization. In order to keep the consistency with the KM approach and for the general applicability of system modelling, UML as the notational standard for object oriented modelling in industries is used to model our KM system. The system modelling involves the steps of use case modelling, class modelling, and the dynamic modelling as discussed below.

4.4.1. System use case modelling

Use case is used to determine the system behaviours before undertaking the detailed system design. It is a common method to describe the interactions between users and the system and to transform the business processes into software systems. In our system, the use cases are built through the following steps: to identify the business processes, to identify the actors, and to create the use case diagrams.

1) Identifying the designer's KM activities

KM activities are the main business processes in our system. Based on the integrated KM approach, designers perform two core KM activities in order to innovate, which are the creation of new knowledge and the creative usage of existing and new knowledge. The use case of knowledge creation depends on several sub use cases: self-learning, modification, creation from existed and creation of really new knowledge. Meanwhile, the use case of knowledge usage contains combining, identifying, commenting, and evaluating knowledge. At the same time the use cases of knowledge creation and usage are supported by knowledge agent who provides and performs other supporting KM activities such as knowledge search, storage, deletion etc. Due to the distributed nature of knowledge, several actors are involved in the creation and usage of knowledge.

2) Identifying actors for use cases

The actors of a use case originate from the artefacts, human and technologies involved in design. They can be classified into business actors and system actors [3]. In our system, business actors are composed by the personal knowledge bases, shared knowledge base, and other ICT tools. System actors are the designers who use the system functions to perform their design activities by creating and using knowledge. A designer can act both as a knowledge creator and as a knowledge user in terms of their activities. Also he can publish his knowledge and access others' knowledge in the shared knowledge base.

3) Creating use case diagrams

For a clear understanding of the interactions between the system and its users, several use case diagrams are proposed with reference to the functions of the system and the actors identified. Figure 6 shows a use case diagram of a designer working as a knowledge creator and user. This diagram illustrates how a designer can create and use knowledge elements for his design tasks when using our system. Others use cases such as sharing and publishing knowledge are not detailed here.

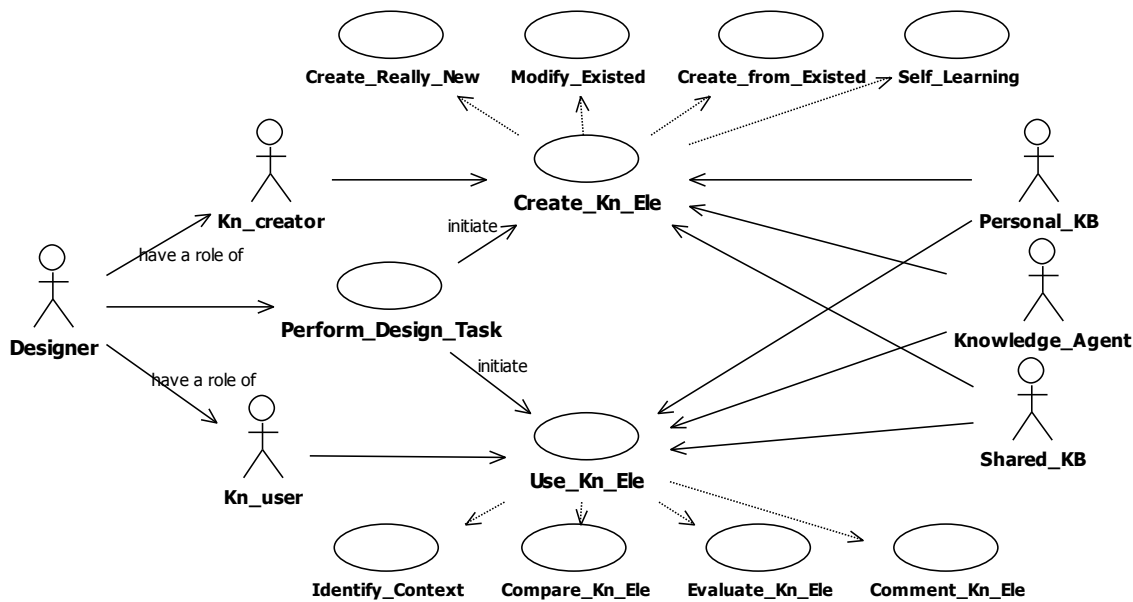

Figure 6 Use case diagram for designer as knowledge creator and user

4.4.2. Class diagram modelling

As we discussed that use case diagrams provides a global and external view of the system, the class modelling is applied for detailing the internal structure of the system by illustrating the objects and their relationships. The class diagrams can help us to inspect the static, relational and structural aspects of the system framework. They can help to improve the understanding of the real structure of KM system and thus provide a sound basis for system implementation. The class modelling mainly contains two steps as follows.

1) Exploring classes of the KM system

Classes of our KM system are generally identified from three main aspects. The first is the business process aspect such as project management and design process; the second is the system organization aspect that is the system framework; and the third is the aspect of use cases. Classes of business processes are the artefacts, human and technologies used in the innovation project and the design tasks in design process. The system framework provides the classes representing the KM activities in the knowledge lifecycle and the existing infrastructures such as the CAD/CAE system, PLM system, database management system etc. The use cases provide the classes that describe the operations of an object, and external entities or actors involved in the system.

The classes identified from three aspects are grouped into several packages according to their functionality in our system. They are the PLM Interoperation Package, Agent Management Package, Knowledge Element Package and Existing Infrastructure Package. For instance, the Knowledge Element Package consists of the classes of content and context model of knowledge element, their relationships and the graphic interfaces for managing it. Agent Management Package contains the classes concerning managing various agent and its working situations. PLM Interoperation Package is composed by the classes that can cooperate with existing information systems. Existing Infrastructure Package is made up of the classes representing the Operating system, CAD/CAE system, DBM system and other involved systems.

2) Creating class diagrams

Class diagram modelling copes with the internal operations of the system. The attributes and operations of classes and the services provided by classes are analysed and detailed in the descriptions of use cases. Also, system functionalities and its framework reflect the required structural relationships between classes, which can be refined to more detailed generation, associations, links or dependencies. Figure 7 shows the complete class diagram of our system built on our integrated approach of KM. In reason of clarity of the figure, the attributes of classes are hided and only the operations and relationships among them are kept.

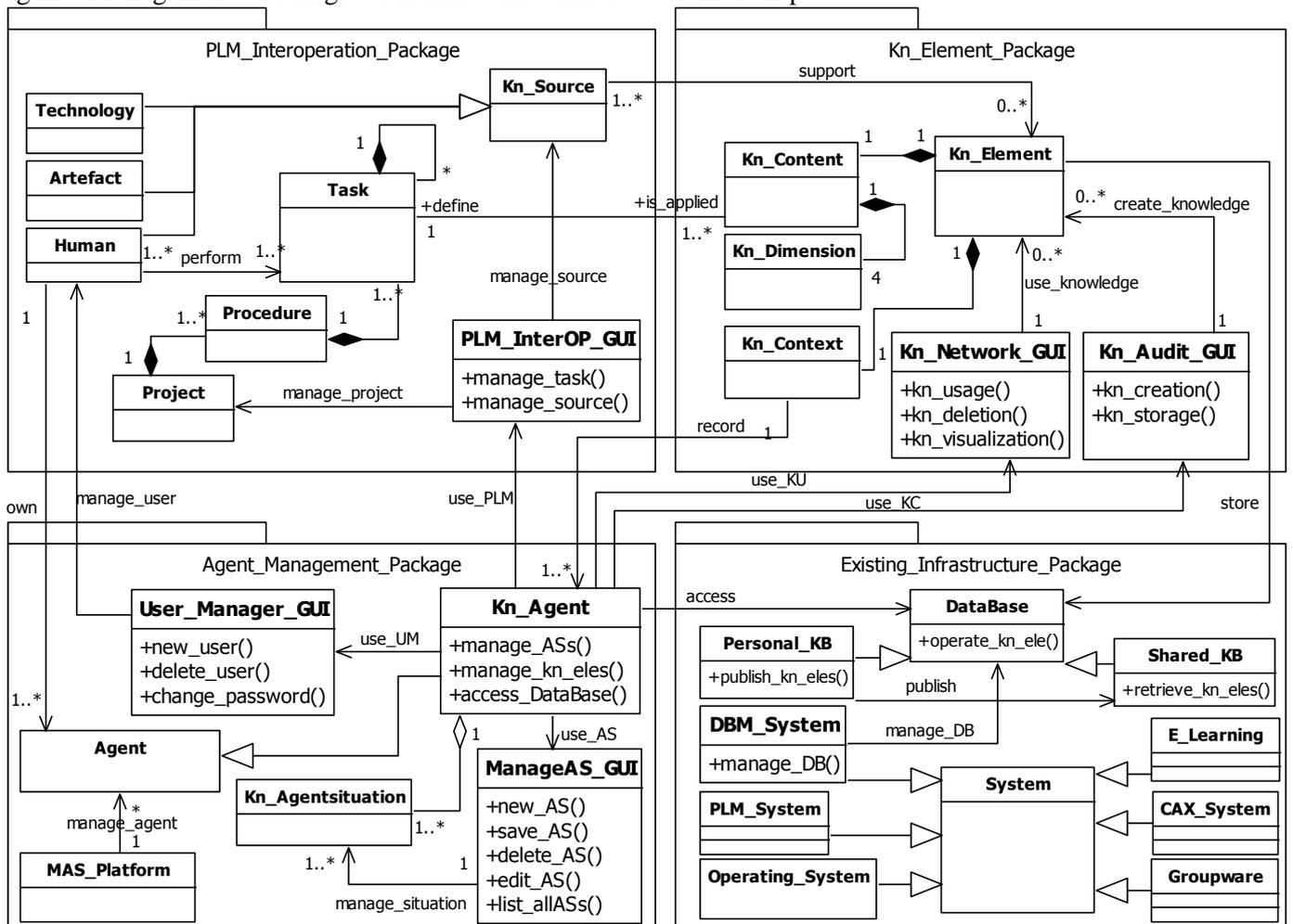

Figure 7 Class diagram of the KM system

Taking the Knowledge Element Package for example, it includes the classes of Kn_Element, Kn_Content, Kn_Context, Kn_Dimension, Kn_Network_GUI, and Kn_Audit_GUI and their internal and external relationships. Kn_Network_GUI represents the interface to use knowledge elements and Kn_Audit_GUI is the interface to create knowledge elements. Kn_Element represents the knowledge model that is composed of Kn_Content and Kn_Context. Kn_Content consists of four dimensions. Each dimension is inherited from the

Kn_Dimension class. Kn_Context can record the contextual information of the knowledge when it is created and used. The detailed explanations about class diagrams in other packages will not be expanded here.

4.4.3. Dynamic modelling of distributed KM system

Due to the dynamic aspect of knowledge and the ever-changing environment, a KM system is far more than a static system. It has extensive interactions with human beings and external environment. Its dynamic behaviours can not be well explained by the static models. Dynamic modelling is an effective way to tackle the dynamic behaviours of the classes in the system. During various time periods, different events and state transitions of an instance are contained in the dynamic models. Sequence diagram and state diagram are very often used to depict the procedural operations and the dynamic behaviours of our system. In the following, the knowledge creation sequence diagram and a design state transition diagram are presented to illustrate system dynamic behaviours.

1) Sequence diagram analysis

A scenario is a sequence of particular events that occur during the execution of a system [3]. Sequence diagram is an important way to represent the causal orders of a series of events and operations described in use cases. The sequence diagram expresses the interactions of associated instances, which are triggered by the stimulus exchanged between these instances. Figure 8 illustrates the sequence diagram of knowledge creation.

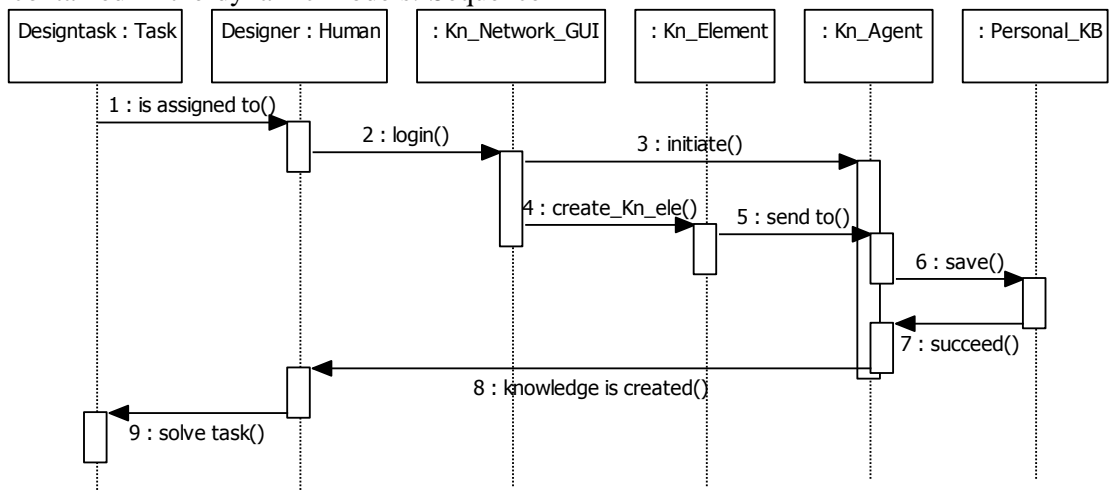

Figure 8 Sequence diagram of knowledge creation

Knowledge creation and usage hold core importance for innovation in design, which are fulfilled by designers with the help of personal knowledge agent. Other sequence diagrams of publishing and sharing knowledge elements and collaborating with other agents are also built for detailing the execution of our system.

2) State diagram modelling

Since the sequence diagrams identify the events taking place among different instances, these events can be used to define the state transitions of instances. A state chart diagram describes the dynamic behaviours of a specific object through the states and transitions. From the point of view of KM, a design process consists of various tasks that undergo a series of state transitions propelled by the accumulation of engineering knowledge. Thus, the state transition of a design task can be modelled based on the integrated approach of KM as illustrated in figure 9.

The transition begins with a search in knowledge bases according the inputs of the design task. Then according to whether there is available knowledge for the task, knowledge use and creation happens respectively. If the task is complicate and it can only be partially solved, then an integration of the partial solutions is introduced. Finally if the task is totally solved, the state is transited in the next state. If not, it will lead to a reformulation of the not solved design task and repeat the above process until a satisfying solution of the task is achieved. With the support of the cross-functional design team and the advanced ICT tools, the state transition model highlights the iterations among states and diminishes unnecessary iterations in the process.

Based on the integrated KM approach, the functionalities and framework of our KM system are defined in this section. The detailed system components are described by the computational models in UML, which build a solid base for the implementation of our system.

## 5. System implementation

Based on the system framework and system modelling in the section 4, the information and computational models of the KM system provide formal descriptions of the realization of our system. In the following, these models are mapped into actual implementation of a software prototype with a platform independent programming language Java. Thus, a distributed KM System for Innovation (abbreviated for "KoSI") has been developed and implemented in the windows operating system by our research group.

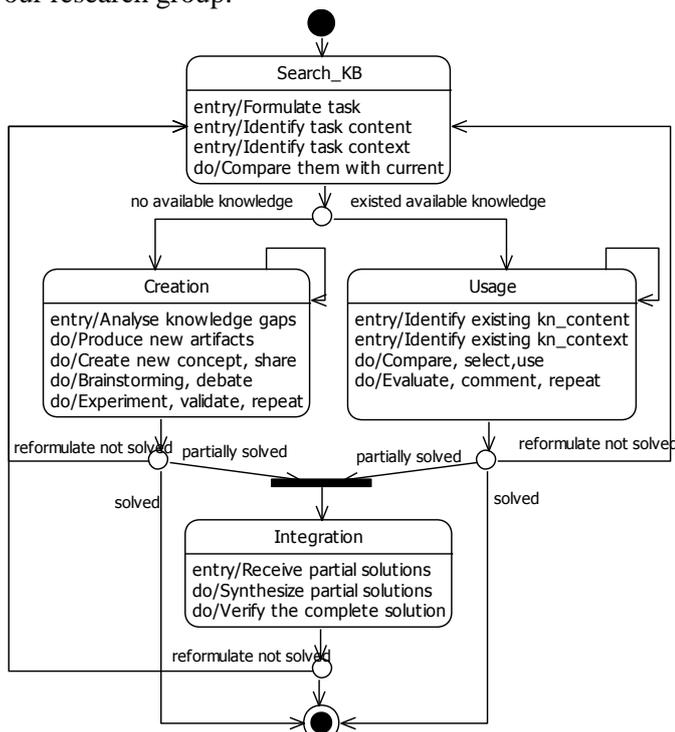

Figure 9 State chart diagram of design state transition

## 5.1. Specifications of KoSI

After we have investigated the current problems of existing KM systems and innovation supporting systems, we conclude that both of the extant systems stand for a part of the requirements of innovation in design through the point of view of KM. According to the C–K design theory in [7, 27], both knowledge and concepts are key actors for innovation. Both of them should be involved in the system through a compatible manner. The systemic model of knowledge and its computational model provide a bridge to integrate them. Since knowledge is the essential asset for innovation, the integrated approach of KM is applied in our system in order to relieve the limitations of current computer supporting systems.

Our system is an independent application and built on the existing information infrastructures such as the CAX system, PLM system and web portal in a company. In terms of the functional requirements and the system framework, the specifications of our system are put forward as follows.

1. Specifications from user point of view:
   ■ A graphic user interface for human – machine interaction is built by the object oriented and agent technologies;
   ■ Personal knowledge bases are decentralized for keeping the privacy of personal knowledge, while a shared knowledge base is centralized for the global access to all stakeholders;
   ■ The collaboration and integration with existing infrastructure are built into the system.
2. Specifications from technological point of view:
   ■ A knowledge template based on C–K theory is created in the XML format based on the systemic model of knowledge, in which the security control and version management are incorporated;
   ■ Knowledge accumulation as the nature of a company becomes a necessary footstone for innovation, so a hierarchical tree mode of knowledge elements depicts the knowledge accumulation mode of innovation;
   ■ As new knowledge and ideas often come from deliberated association and combination of the incompatible objects in a creative manner, a networking mode of knowledge elements with interrelationships and dependencies represents the knowledge connectivist mode of innovation.

To ensure the robustness and mobility of our system in a distributed environment, the object-oriented technology and Java language are adopted for the high compatibility. For keeping the consistence with industrial standard object models and the interoperability with other engineering information systems, the UML models are translated into the computer models and programming codes for the development of KoSI.

## 5.2. System architecture selection

The system architecture is a computer presentation of system frameworks, where the system packages and components are arranged on a hierarchy of layers with well-defined interfaces. Currently, the most diffused and mature architecture for realizing distributed systems is client/server architecture, whereas it has its inherent deficiencies such as the rigid distinction between client and server, the centralization and reactivity of server and the inability of client.

In the distributed context, different people have various attitudes and points of view towards KM and their knowledge is also heterogeneous and distributed. The agent paradigm has been seen as an appropriate way to cope with the heterogeneity, diversity and flexibility of KM systems [34]. By considering the specifications and the framework of our system, client/server architecture can efficiently offer a shared and centralized knowledge base that is an important component and functional service for KM activities. Thus, taking account of the security and maturity of client/server structure and the flexibility of the agent technology paradigm, the hybrid system architecture is suitable and is adopted for KoSI.

The hybrid structure is a combination of client/server structure with distributed agent paradigm. The agent at each client end could assist the owner's KM activities according to his personal wishes and objectives. The server provides a common place to share the knowledge and collaborate among agents and individuals. The application logics concerning individual activities are built on the client end, while the common rules about collaboration and sharing are defined at the server end. With this architecture, the information traffic cost, security and flexibility can well supported and guaranteed.

## 5.3. Presentation of the interfaces of KoSI

Based on analyses of the system specifications and its architecture, KoSI is developed and implemented at our laboratory and the partner company. Initially, it is realized in a portable way, where the client and server are located at the same machine with a planned expandability into the real distributed system. The interfaces of KoSI are presented in the following.

The main frame of our system is a single application framework, where the function modules such as system login and user management are incorporated as shown in figure 10-A & 10-B. Login interface is responsible for entry control of the system; user management interface provides an access to organize all the users in the system. The two core functional components are the interfaces of knowledge creation and usage that are Knowledge Audit and Knowledge Network interfaces. The knowledge Audit interface is shown in figure 11-A, where new knowledge is created in a template with the creator's context automatically captured. The Knowledge Network interface contains a tree view and a networking view of knowledge elements in the knowledge repositories according to the criteria selected as in figure 11-B.

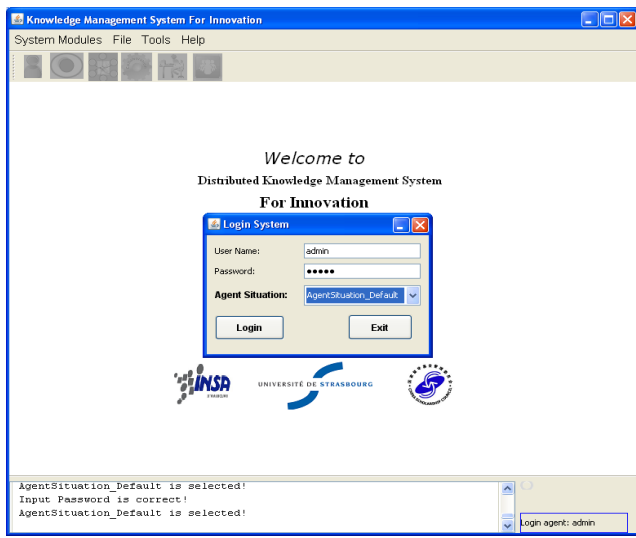 (A) 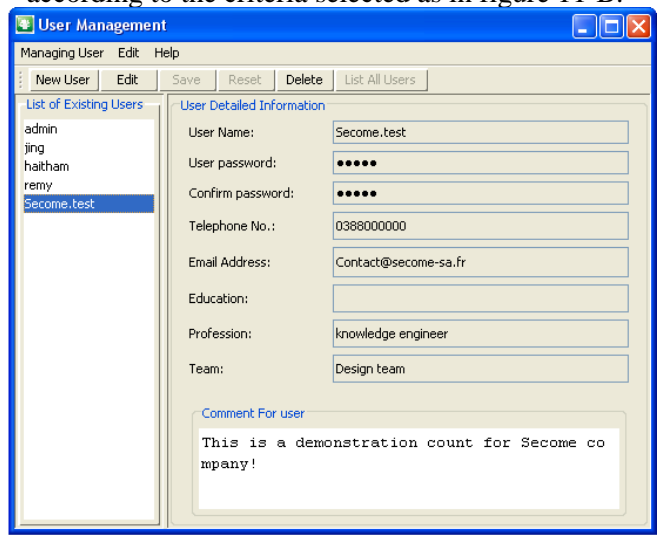 (B)

Figure 10 Login and user management interface

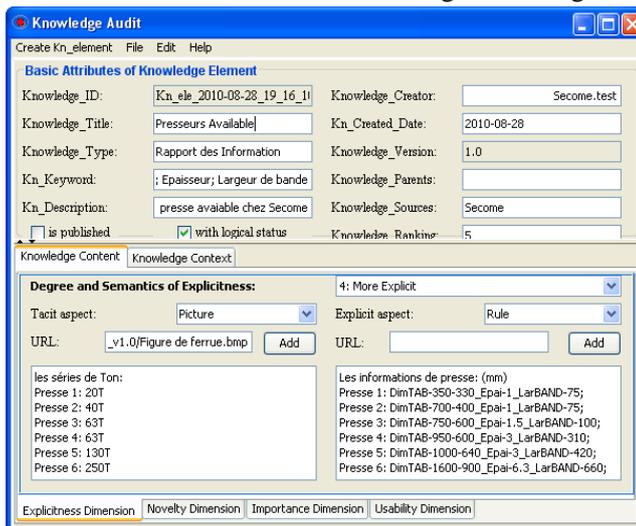 (A) 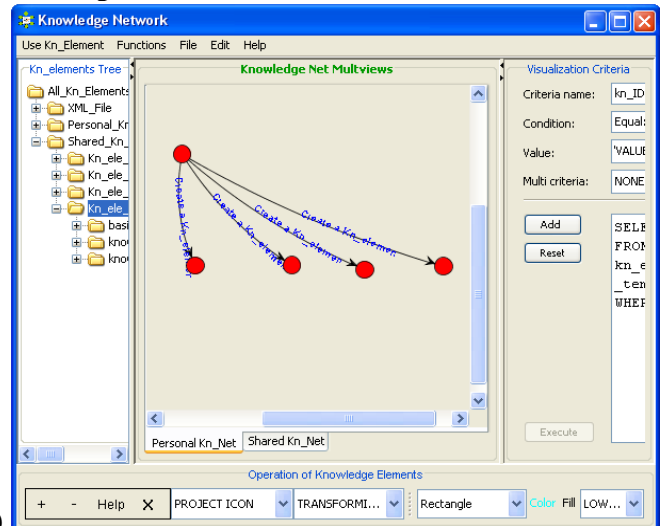 (B)

Figure 11 Knowledge creation and usage interfaces

In terms of the integrated approach of KM, the two core activities: knowledge creation and usage are supported by other KM activities such as the search, storage and navigation of knowledge. These supporting functions of KM in the Pre-creation, Intermediate and Post-usage phases are executed in the background and intensively facilitated by the knowledge agent and ICT tools.

In reality, a user performs various tasks at the same time and creates and uses his knowledge so as to fulfil them. Thus, an interface of PLM interoperation is created for managing the tasks, procedures and projects of a user. In order to collaborate with existing system infrastructure in a company, task information can be imported and exported in the XML format. Since personal knowledge agent helps a user to organize his tasks and responds to the user requirements of knowledge, an interface of agent management is created to manage different agents and their working situations. Other detailed interfaces for security control and change of working situation will not be explicated here.

Based on the integrated KM approach, the interfaces of Knowledge Audit and Knowledge Network correspond to the two core KM activities for innovation and are situated at the centre of KoSI prototype system. Other interfaces such as task and agent management are created for supporting their functioning in the system. The KoSI prototype demonstrates the functionality and validity of the proposed integrated KM approach. The interfaces in the prototype provide a convenient way for designers to create new knowledge and use it more efficiently so that their creativity can be greatly released for more innovation in design. The KoSI prototype has been recently applied in our partner company in a project of innovative die design for sheet metal production. The initial results are introduced in the following section 6.

## 6. Application and qualitative evaluation

Tooling design is a knowledge-intensive process where much tacit and imprecise knowledge is held and applied by design experts [35]. In order to facilitate the innovation in tooling design, experts' knowledge needs to be capitalized and shared for more creative usage and creation of knowledge. Our prototype has been accepted and applied in a project of tooling design in our partner company. The application is presented in the following and its initial results are evaluated qualitatively with reference to a design research approach and the criteria of evaluation in [36]. The quantitative evaluation has been planed as the object of future work.

6.1 An industrial case study

Our partner is a small and medium company who specializes in the design of progressive die for sheet metal production. As soon as a client sends an order of a sheet metal part together with its engineering documents, the company will propose a technical solution for the order, which contains a progressive die design for the part. In our application, the progressive die is chosen as our research object. Figure 12 indicates the engineering model of the part.

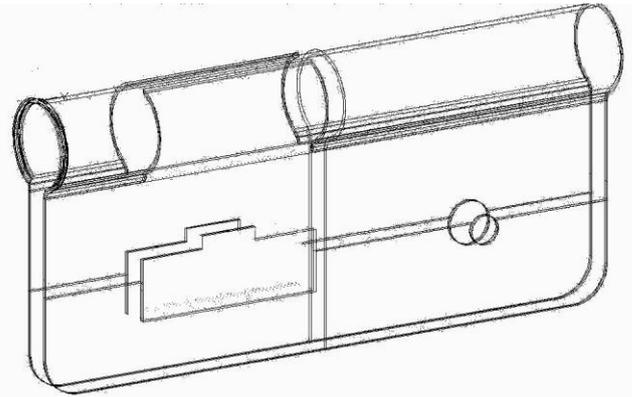

Figure 12 Engineering model of sheet metal part

After studying the existing design projects of similar parts, a scenario of typical die design process is proposed for designing this new progressive die. It begins with a reception of order, then the pre-study of the feasibility. And then it is followed by unfolding the part, estimating the dimensions of the die and creating the technical solution. After, with the negotiation with the client, the order could be accepted, reconsidered or refused. If it is accepted, a detailed study of the technical solution will be performed. This scenario is illustrated in figure 13.

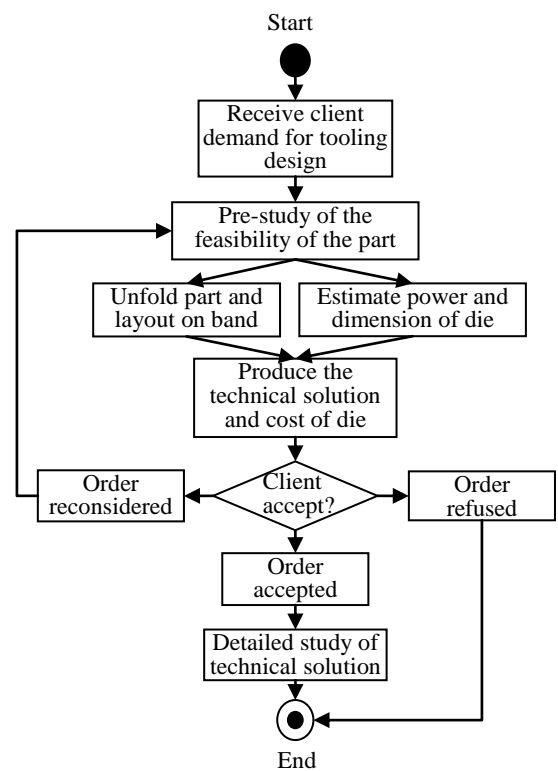

Figure 13 Scenario of die design for sheet metal

In the application of KoSI prototype, as soon as an order from a client arrives with its technical documents, a design project and a working situation are created in the system. The die designer retrieves existing knowledge elements about similar cases from knowledge bases and pre-studies the feasibility of the new part. The available knowledge elements are visualized in a network with relation to properties of the new part. According to their relevance, some elements are selected as references for studying of feasibility of the new part in figure 14.

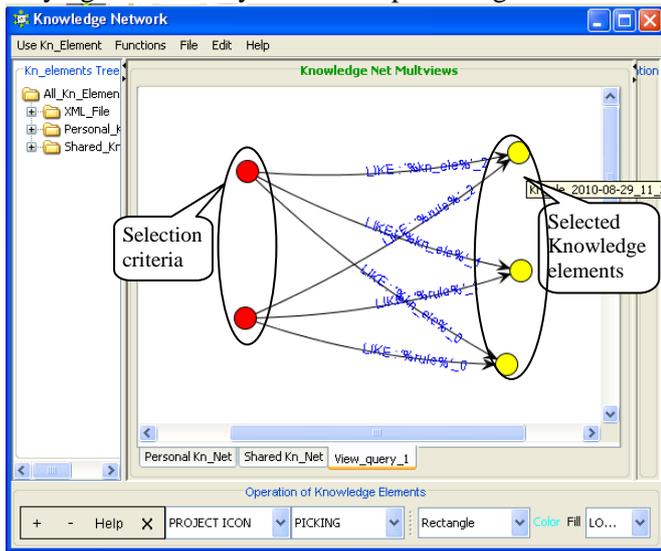

Figure 14 Network of selected knowledge elements

According to the existed knowledge about feasibility and designer's experience, the new part is considered as feasible with current production facilities. This leads to further steps of unfolding the part and estimating the power and dimensions of the die. With reference to the structure of the part, a rolling feature has been noticed, which is difficult to realize and critical for the design of its die. To create the feature, two movements should be fulfilled and coordinated in order to obtain the right form. They are the vertical movement of the press and the rolling movement of the part.

According to the designer's experience, a new neutral line is defined on the unfolded part and new forming and rolling steps are designed in order to coordinate the two movements. Thus, two new knowledge elements are created in the XML format and stored in the knowledge base, which concern the definition of a new neutral line on the unfolded part and the arrangement of forming steps as explained in figure 15-A. Meanwhile, a lot of knowledge elements are shared and provided by other team members so that useful information is available for completing the layout design of the progressive die as shown in figure 15-B.

(A) 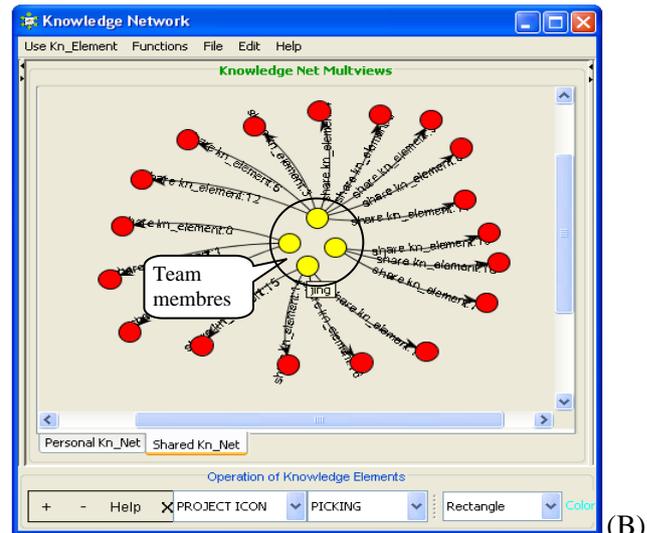 (B)

Figure 15 New and shared knowledge elements

The new knowledge elements correspond to the innovative solutions in the layout design of progressive die. The neutral line of the unfolded part is defined as the direction of band feeding and the centre line of its positioning, which ensure the continuity of the production process. The new arrangement of forming and rolling steps helps to relieve the complexity of die and to create a more balanced layout. The final innovative layout of the progressive die is illustrated in figure 16. The fourth and fifth steps in the layout innovatively resolve the quality problem of forming of the part while keeping the successiveness of production.

6.2 Qualitative evaluation

Assessing the performance of the integrated approach of KM is an important as well as a difficult task. It is partially because of the difficulty to establish the metrics to evaluate the improvements of design solutions and to assess the long term effects of implementing a KM system in a company. Thus, in order to reduce the difficulty, we perform a qualitative evaluation of our application and then plan the quantitative evaluation as future work.

In terms of the criteria of evaluation in [36], we identify three indicators for qualitatively evaluating the efficiency of our prototype in the application. The indicators are the performance of design solution, the time of development and the return of investment. According to the initial results of the application, we have qualitatively observed several benefits of our prototype system in this application, which include:

1) the automatically capturing the design expert's working context and integrating it into the knowledge base;
2) the provision of multiple methods for the expert knowledge representation;
3) the tree mode and network mode of the visualization of knowledge elements facilitating the usage and creation of knowledge for innovation;
4) the improved traceability and trustworthiness of the knowledge elements, which lead to better decision making in design.

The prototype is currently applied by die designers in the company. With the new knowledge created and used in the progressive die design, the performance of design solution is qualitatively improved due to the continuity of the production and a more balanced layout of die. As our prototype provides convenient ways to find related knowledge and to use them in design, the time of development of the die is also significantly reduced. However, due to the long term nature of the return of investment, we can only estimate that our prototype has positive effects on this indicator.

Even though we have obtained some initial results of our application, they are far from perfect. As the evaluation system of the indicators is still not well established in our partner, more work needs to be done to quantitatively evaluate our approach. Further applications and research have been planed to continuously improve the evaluation of the application of our approach and the prototype.

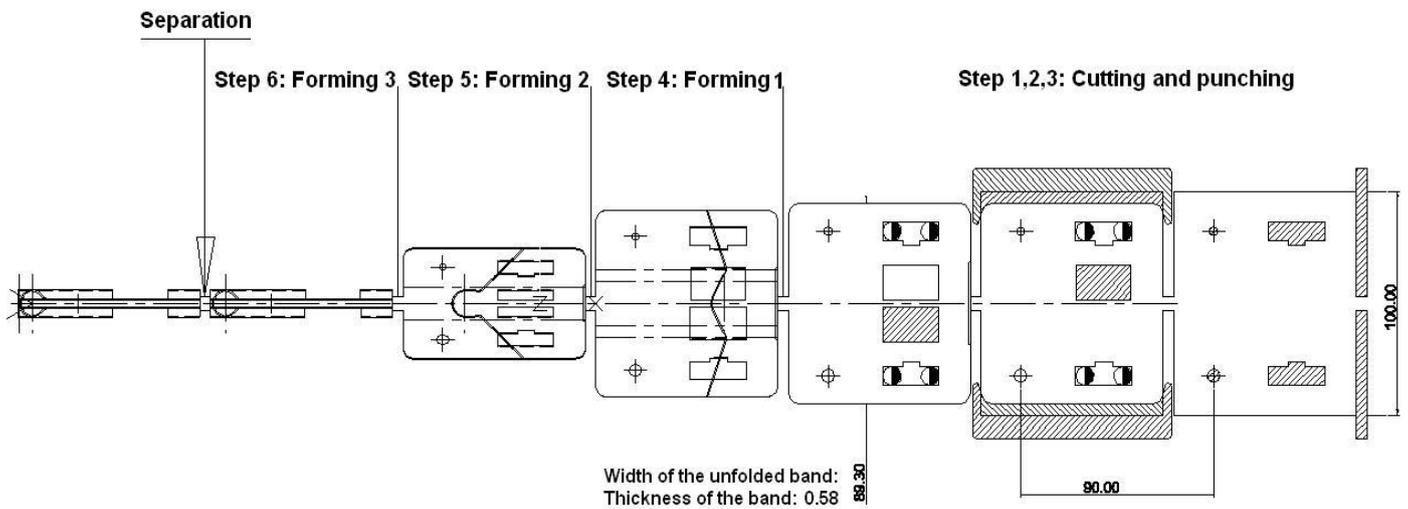

Figure 16 Innovative layout of the progressive die

## 7. Conclusion and perspective

As innovation has been regarded as an imperative for the success of a company, how to leverage the available knowledge into real innovation in the market is the focus of our study. In order to understand the problem, KM processes, innovation models and design methods are exploited to explore the complex relationships between KM and innovation. The contribution of our work is the description of an integrated approach of KM to foster innovation in design. The approach rests on a systemic model of knowledge and a hierarchical model composed by the macro process and meta-model of KM. They provide an effective means of integrating different perspectives of KM and leveraging KM activities into innovation. And thus, it enables the designers to innovate more easily and efficiently in a knowledge-intensive and dynamic environment. With the integrated KM approach, design knowledge can be efficiently created and used and its availability is greatly enhanced so that the complexity and uncertainty of innovation are reduced.

Based on this approach of KM, a distributed knowledge management system for innovation is modelled and realized in order to relieve some disadvantages of current KM and innovation supporting systems. The prototype of our system demonstrates the functionalities and validity of the integrated approach. Finally, the application of our prototype in an innovation project of progressive die design is introduced and exhibits the applicability and feasibility of our approach to help designers to innovate.

Although substantial endeavours have been made in this study to develop the integrated approach and the prototype for fostering innovation in design, there are still several issues to be addressed in further research. Due to the time span of evaluation, only the qualitative evaluation is presented here. Practically, further applications and case studies are necessary and planed in order to fully establish the evaluation system and to quantify the indicators for more detailed evaluation of our approach. The customization of the prototype and its transition to web-based and distributed environment will be further researched.

## Acknowledgements

We are grateful to the Chinese Scholarship Council (CSC) and our partner company for their support of our research.